\documentclass[twocolumn,amsmath,amssymb,showkeys]{revtex4}

\usepackage[latin1]{inputenc}
\usepackage{color}
\usepackage{graphicx}
\usepackage{subfig}
\usepackage{latexsym}
\usepackage{natbib}
\usepackage{mathtools}
\usepackage{hyperref}

\newcommand{\ep}{\epsilon}
\newcommand{\ga}{\gamma}

\newcommand{\ka}{\kappa}

\newcommand{\La}{\Lambda}

\newcommand{\Om}{\Omega}
\newcommand{\om}{\omega}
\newcommand{\si}{\sigma}

\newcommand{\be}{\begin{equation}} 
\newcommand{\ee}{\end{equation}}
\newcommand{\bea}{\begin{eqnarray}} 
\newcommand{\eea}{\end{eqnarray}}

\begin{document}
\title{An analytic approach to baryon acoustic oscillations}
\author{Francesco Montanari}
\email{francesco.montanari@oamp.fr}
\author{Ruth Durrer}
\email{ruth.durrer@unige.ch}
\affiliation{D\'epartement de Physique Th\'eorique, Universit\'e de Gen\`eve, 
24 Quai E. Ansermet, 1211 Gen\`eve 4, Switzerland}
\date{\today}

\begin{abstract}
The fitting formula for the location of the first acoustic peak in the matter power 
spectrum is revised. We discuss the physics that leads to baryon acoustic oscillations: the recombination history, the tight coupling approximation and the velocity overshoot effect.
A new fitting formula is proposed, which is in accordance within $5\%$ with numerical results
for a suitable range of cosmological parameters, whereas previous results  yield deviations
of up to $(15-20)\%$.
The crucial improvement turns out to be the accuracy of the recombination history.
\end{abstract}

\maketitle

\section{Introduction}
Before recombination the Universe is dense and highly ionized, and baryons and photons 
are tightly coupled by Thomson scattering.
The pressure of the Cosmic Microwave Background (CMB) photons opposes gravitational
collapse and leads to acoustic oscillations.
In fact, during this phase, the amplitude of perturbations in the baryon
density cannot grow, but they oscillate with a slowly decaying amplitude.
After recombination, baryons decouple from radiation and the oscillations are `frozen in'.
Because baryons represent a significant fraction of matter, cosmological perturbation
theory \citep{HS1996} predicts that these acoustic oscillations of the baryons 
(BAO's) are imprinted on the late-time matter power spectrum, leaving features
analogous to the acoustic peaks in the CMB power spectrum.
The BAO's have indeed been observed in the large scale galaxy 
distribution~\citep{eisenstein:2005,tian,arnalte-mur}, and they are one of the main 
observational goals of recent and upcoming surveys.

Numerical calculations  of the recombination history are available thanks to, e.g., the {\sc Rico} \cite{Fendt:2008uu} or the {\sc Recfast} \cite{Seager:1999New} codes for the recombination history. The latter, in particular, is the default model used in the {\sc Camb}
Boltzmann-solver code \citep{Lewis:1999bs}.
It does reproduce the results shown in~\cite{Seager:1999New} and is a fast 
approximation to the detailed calculations described in \cite{Seager:1999km} 
with some updates discussed in \cite{Wong:2007ym} and with the Compton 
coupling treatment of \cite{Scott:2009}.
The numerical calculation of the recombination history is much more time expensive 
than employing analytical approximations or fitting formulae, like the ones used 
by~\cite{EH98} which are an improvement of the fits presented in Appendix~D 
of~\cite{HS1996}.
Nevertheless, an accurate computation of the recombination history turns out to be a significant step for the evaluation of the location of troughs and peaks in the transfer function.

At early times, before recombination, baryons and photons behave as a single `tightly 
coupled' fluid because Thomson scattering, which couples electrons and photons, 
is much more rapid than the expansion of the universe $t_s \ll H^{-1}$, where 
$t_s=(\si_Tn_e)^{-1}$ is the photon scattering time scale (i.e. the mean time between 
Thomson scatterings) and $H^{-1}$ the expansion time scale of the universe. Here $\si_T$ 
is the Thomson cross section and $n_e$ is the number density of free electrons.
Since scattering is rapid compared with the travel time across a wavelength, 
we can expand the perturbation equations in powers of the Thomson mean 
free path $\lambda_s=t_s = \dot{\kappa}^{-1}$ over a wavelength 
$\lambda\propto k^{-1}$, i.e. $k/\dot{\kappa}$, where 
$\dot{\kappa} = a\si_Tn_e$ is the differential optical depth.
To the lowest order we obtain the tight coupling approximation (TCA) \citep{MaBert95}.
A more rigorous definition and treatment of the TCA can be found in \cite{pitrou}, while \cite{cyr-racine} analyzes the second-order approximation in the inverse Thomson opacity expansion.
In \cite{Blas:2011rf} formulas for the TCA to second-order are derived independently and tested. 
These results are implemented by the {\sc Class} Boltzmann code.

It can be demonstrated that at large scales the transfer function is governed by density perturbations, which oscillate roughly like a $\cos\left( ks\right)$ where $s$ is the sound 
horizon at decoupling, see e.g. \cite{HS1995} and the Appendix D of \cite{HS1996}.
However, the corresponding velocity perturbation dominates in the small scale limit.
When the oscillations are released at decoupling,  baryons move kinematically according 
to their velocity and generate a new density perturbation \citep{EH98}.
This `velocity overshoot' effect is responsible for the fact that the transfer function, 
for sufficiently large $ks$, actually behaves like $\sin\left( ks\right)$.

In this paper we derive a fitting formula for the location of the peaks and troughs in the matter power spectrum  by matching the solutions for the matter density perturbation before and 
after decoupling. We obtain a form which is consistent with the one proposed in \cite{EH98} 
for the position of the first peak.
However, our fit is tested considering recent cosmological parameters and 
it uses an improved recombination history.
The latter turns out to be an important amelioration which lets us achieve a significantly 
better accuracy than the previous fit.
Even though one can compute the positions of these peaks numerically with the help e.g. of 
{\sc Camb}-code, we believe that an analytical fit has its merits as it helps us to see immediately
what effect a variation of cosmological parameters will have and since it gives us a better
understanding of the physics involved.

The paper is organized as follows.
In Section~\ref{s:BAO} we review the physics leading to BAO: the ionization history, the TCA 
and the velocity overshoot effect.
In Section~\ref{s:1pk} we give the fitting formula and we compare the results with \cite{EH98}.
We state our conclusions in Section~\ref{s:concl}.
Some details of the derivations are moved to an Appendix A.

\section{Baryon Acoustic Oscillations}
\label{s:BAO}

We review the physics leading to BAO, in order to highlight the main results that allow us to 
locate the peaks and troughs in the transfer function.

We use the notation of~\cite{durrerPT}. In particular, $t$ denotes the cosmic time and 
$\eta$ conformal time such that $ad\eta = dt$, where $a$ is the scale factor. An over-dot 
indicates a derivative w.r.t. the conformal time $\left(\ \dot{}\ \right)\equiv d / d\eta\left( \right)$.
We also use of the notation $R\equiv 3\rho _b / 4\rho _{\gamma}$, where the 
subscripts $b$ and $\ga$ label the energy density of baryons and photons, respectively.
Our reference model is a $\La$CDM Universe.

\subsection{Ionization history}
In \cite{Seager:1999New} a calculation of the recombination of H, He I, and He II in the early Universe is developed which is implemented in the publicly available code {\sc Recfast}.
The methodology is to calculate recombination with as few approximations as possible.
One of the main improvements with respect to previous calculations is that it takes into
account that the  population of excited atomic level depart from an equilibrium distribution.
Indeed, recombination is not an equilibrium process.
Simplified analytical calculations or approximate fitting formulae for the 
recombination history are too crude to give good approximations for the location of peaks 
and troughs in the matter transfer function as we will discuss in \S\ref{s:1pk}.

{\sc Recfast} models hydrogen and helium atoms as effective three-level systems, including some artificial corrections via conveniently chosen \emph{fudge factors} \cite{Wong:2007ym}.
An alternative approach is the {\sc Rico} code \cite{Fendt:2008uu}, which smoothly interpolates the ionization fraction on a set of pre-computed recombination histories for different cosmologies.
The inclusion of previously neglected physics leads to changes in the ionization fraction at the $(2 -  3\%$ level in some redshift regions.
However, throughout this paper, we run the {\sc Camb} code using {\sc Recfast} version {\sc 1.5}, which gives a sufficiently accurate recombination history for our purposes.

On the other hand, we treat the late-time reionization of hydrogen and helium via the fitting formulae proposed in the appendix of \citep{Lewis:2008wr}.

\subsection{Tight Coupling Approximation}
Here we derive an analytical solution for the baryon density perturbation in the tight coupling approximation (TCA) in first order perturbation theory using the WKB approximation valid 
for a slowly varying $R$,  inside the sound horizon at decoupling given by
$s\simeq c_s^{(\gamma b)}\eta_{dec}$. Here $c_s^{(\gamma b)}$ is the sound speed of the baryon-photons fluid (defined in appendix \ref{app:TCA}, Eq.~(\ref{eq:c2s_gr})), and $\eta_{dec}$ is the decoupling time.
Hence, we must keep in mind that this approximation is valid only for sufficiently large $k$.
For the range of our interest this is fine since acoustic oscillations concern relatively 
small scales, of the order of $100h^{-1}$ Mpc \citep{eisenstein:2005}.

We perform our calculation in the uniform curvature gauge, the differences between 
the variables calculated in different gauges is small on sub-horizon scales \citep{durrer}.
Furthermore, all the physical observables must be indeed gauge invariant.
So, in terms of the density perturbation in the uniform curvature gauge, $D_g$ \citep{durrerPT}, the general tight coupling solution for the baryon density perturbation is given by
(see appendix \ref{app:TCA} for a derivation)
\begin{subequations}
\begin{eqnarray} \label{eq:bar_an}
D_{gb}^{(t.c.)} \left(k, \eta\right) &=& D_{gb}^{(in)} \left(\frac{1}{1+R\left(\eta\right)}\right)^{1/4} \cos\left( kr_s\right) \nonumber \\
&& - E\left( k,\eta \right) \;,
\end{eqnarray}
where
\begin{eqnarray} \label{eq:bar_an_int}
E\left( k,\eta \right) &=& \left( 1+R\left(\eta\right)\right)^{-1/4}  
\int _0^{ \eta}d\zeta\ \left[ \frac{2+R\left( \zeta \right)}{\left(1+R\left(\zeta\right)\right)^{3/4}} \right. \nonumber \\
&& \times \left. \frac{\sin\left[kr_s\left( \eta\right)-kr_s\left( \zeta \right)\right]}{kc^{\left( \gamma b \right)}_s\left(\zeta\right)} k^2\Psi\left( k,\zeta\right) \right] \;.
\end{eqnarray}
\end{subequations}
$D_{gb}^{(in)} = (3/4)D_{g\ga}^{(in)}$ is determined by the adiabatic initial condition and 
$\Psi\left(k,\eta\right)$ is the Bardeen potential~\citep{durrer}. We have introduced the 
(comoving) sound horizon $r_s\left(\eta \right) \equiv \int _0^\eta d\zeta 
c^{\gamma b}_s\left(\zeta\right)$, i.e., the distance that a wave can travel in a time $\eta$.
During the tight coupling phase the baryon density perturbations undergoes harmonic motion following roughly a cosine mode with an amplitude that decays in time as 
$\left(1+R\left(\eta\right)\right)^{-1/4}$.

To show that this solution follows a cosine mode, a simple analytical approximation of 
the Bardeen potential $\Psi\left(k,\eta\right)$ can be obtained by writing the Bardeen 
equation in the case of adiabatic perturbations for a mixture of perfect fluids (photons, 
baryons and CDM). On super-horizon and sub-horizon scales one finds~\citep{durrer}, respectively,
\begin{subequations}
\begin{eqnarray}
&& \Psi_{x\ll 1}\left(k,\eta\right) = \Psi _0\left(k\right) \label{eq:PsiLarge} \;, \\
&&\Psi _{x\gg 1}\left(k,\eta\right) = -3\Psi _0\left(k\right)\frac{\cos\left(x\right)}{x^2} \;, \label{eq:PsiSmall}
\end{eqnarray}
\end{subequations}
where the initial metric perturbation $\Psi _0\left(k\right)$ is constant in time and 
$x\equiv k\int_0^\eta c_s^{(\gamma b)} d\eta$.
To derive these relations, we also assume $c^2_s\sim \left( c_s^{(\gamma b)}\right)^2\simeq 
1/3$. Here $c_s = \dot{P}/\dot{\rho}$, where $P$ and $\rho$ denote the total pressure and 
energy density of the baryon-photon fluid, respectively.
The latter approximation means that, since the WKB approximation requires slowly varying $R$, we suppose $\dot{R}\simeq 0$ over an oscillation period.
This implies that we are also approximating the equality epoch as roughly the decoupling epoch.

Computing the integral in Eq.~(\ref{eq:bar_an_int}) we obtain an analytical approximation for 
the baryon density perturbation in the tight coupling limit.
Neglecting the small contribution from $x<1$ in the integrand of Equation (\ref{eq:bar_an_int}), using Equation (\ref{eq:PsiSmall}) and $R\simeq 0$, we obtain
\begin{subequations}
\be
E(k,\eta) = -6\Psi_0I(x)/(c_s^{{\ga b}})^2 \;,
\ee
with
\begin{eqnarray}
I\left( x\right) &=& \int_1^x \frac{\cos(\xi)\sin(x-\xi)}{\xi^2} d\xi \nonumber \\
&=& \bigg[ - \cos(x) {\rm Ci}(2\xi) - \sin(x) {\rm Si}(2\xi) \nonumber \\
&& \left. \left. - \frac{\cos(\xi)\sin(x-\xi)}{\xi} \right] \right\rvert _{\xi=1}^{x} \;, \label{Ix}
\end{eqnarray}
\end{subequations}
where $\text{Si}\left( \xi \right) \equiv \int _0^\xi d\chi \sin \chi / \chi$ and $\text{Ci}\left(\xi\right) 
\equiv -\int _{\xi}^{\infty} d\chi \cos \chi/\chi$ are the sine and cosine integral functions.
Finally, we consider modes $x\gtrsim \pi$ and use approximately the asymptotic form $\text{Si}\left( x \right) \to \pi/2$ and $\text{Ci}\left( x \right) \to 0$ valid for $x \gg 1$, we obtain
\be
I(x) \simeq \pi ^{-1} \sin (x) \;.
\ee

\begin{figure}[h]
\begin{center}
\includegraphics[width=\columnwidth]{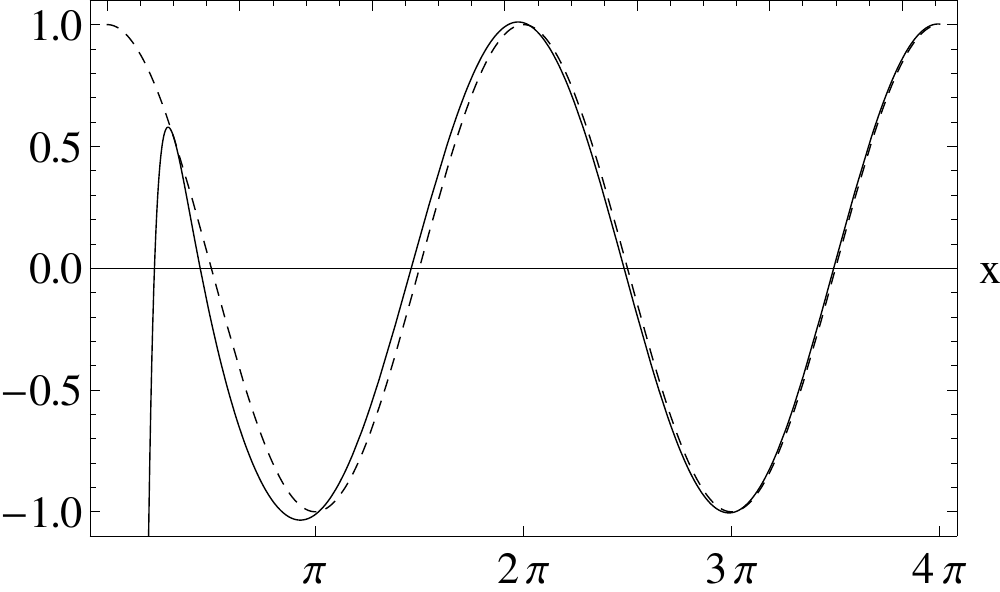}
\end{center}
\caption{The solid line shows Equation (\ref{eq:DbSmall}), with $D_{gb}^{(in)}=1$ and $I(x)$ evaluated as in Eq. (\ref{Ix}).
The dashed line is $\cos x$.
The curves are in good agreement for $x\gtrsim 3\pi$ and 
deviations are significant only for $x\lesssim \pi$.}
\label{fig:DbSmall}
\end{figure}

We summarize our result for the tight coupling approximation in the form
\begin{subequations}
\begin{equation} \label{eq:Db_tc_approx}
D_{gb}^{(t.c.)} \left( k,\eta\right) \simeq D_{gb}^{(in)}\cos\left( kc_s^{\gamma b}\eta\right) - \Psi _0g\left( k,\eta\right) \;,
\end{equation}
where
\begin{equation}
\label{gx}
g\left(x\right) = - \frac{6\; I\left( x\right)}{\left(c_s^{\gamma b}\right)^2} \simeq - 18\, \pi ^{-1} \sin (x) \;. 
\end{equation}
\end{subequations}
%
Making use of the perturbed Einstein constraint equations and of the Friedmann equations to rewrite $\Psi_0$, we obtain 
\begin{equation} \label{eq:DbSmall}
D_{gb}^{(t.c.)} \left(x\right) \simeq D_{gb}^{(in)}\cos\left( x\right) - 12D_{gb}^{(in)}\frac{I\left(x\right)}{x^2} \;.
\end{equation}
Deviations of $D_{gb}^{(t.c.)}$ from the cosine mode decay like $x^{-2}$.
In Fig.~\ref{fig:DbSmall} we compare the pure cosine mode with the full approximate 
solution given by Equation (\ref{eq:DbSmall}).
For $x\gtrsim 3\pi$ the deviation from the cosine mode is negligible, and only for 
$x\lesssim \pi/2$ the integral term is dominant.

In conclusion, the peaks of the tight coupling solution for baryon density perturbations 
$D_{gb}^{(t.c.)} \left( k,\eta\right)$ closely follow those of $\cos\left(kc_s^{\gamma b}\eta\right)$ 
for small scales, say $x \gtrsim 3\pi \simeq 10$, while the few first peaks may exhibit 
deviations due to the integral term.
In particular we expect large deviations from the cosine mode for $x\lesssim 1$.
This consideration allows us to find a formula to fit the position of the peaks and of the troughs of the matter power spectrum.

\subsection{Velocity overshoot}
The velocity overshoot effect can be explained by noting that 
decoupling  is close to equality,  $\eta_{eq}\lesssim \eta_{dec}$.
 Before $\eta_{eq}\sim \eta_{dec}$, baryons are tightly coupled to photons and their 
 velocity is governed by the dynamics of photons, which are the dominant component. 
 Indeed, $V_b=V_{\ga}\leadsto\sin\left( ks\right)$, where  $\leadsto$ indicates 
 `oscillates as', is the tight coupling solution in the limit $\dot{\ka}\to\infty$ (see 
 Appendix~\ref{app:TCA}, Eq.~(\ref{eq:Vb_Vg_TC})).
When $\eta>\eta_{eq}$, the energy density of photons $\rho_{\ga}$ becomes smaller
than the matter energy density $\rho_m$.
Furthermore, for $\eta>\eta_{dec}$ baryons are no longer coupled to radiation.
This implies that for $\eta>\eta_{eq}\sim \eta_{dec}$, baryons no longer follow the photon 
velocity. As we shall see, the baryon velocity after decoupling, $V_b\left( \eta>\eta_{dec} \right)
\leadsto\cos\left( ks\right)$ is almost exactly out of phase with $V_{\ga}\left( \eta>\eta_{dec}\right)
\leadsto\sin\left( ks\right)$.

This can be shown by matching the solutions for the baryon density perturbation before 
and after decoupling.
As derived above, the adiabatic initial conditions for an inflationary model select the cosine 
mode for the baryon density perturbation tight coupling solution on small scales, which is 
given in Eq.~(\ref{eq:Db_tc_approx}) for $kc_s^{(\gamma b)}\eta\gtrsim\pi$.
Let us indicate $D_{gb}$ the solution after decoupling; we match it to the tight 
coupling solution, 
\begin{subequations}
\begin{eqnarray}
&&D_{gb}(k,\eta_{dec}) = D_{gb}^{(t.c.)}(k,\eta_{dec}) \label{eq:match_D} \;, \\
&&\dot{D}_{gb}(k,\eta_{dec}) = \dot{D}_{gb}^{(t.c.)}(k,\eta_{dec}) \;. \label{eq:match_dD}
\end{eqnarray}
\end{subequations}
After decoupling baryons evolve like CDM. The evolution of $D_{gb}$ can then be 
evaluated by considering the Bardeen equation for a mixture of non-interacting radiation 
and matter fluids in a matter dominated epoch that, neglecting the decaying mode, yields 
$\Psi = \Psi(k,\eta_{dec})$ constant in time.
Using equations (\ref{eq:match_D}) and (\ref{eq:match_dD}) as initial conditions and 
denoting the present time by $\eta_0$, on small scales $ks\gg1$ we obtain
\begin{eqnarray}
D_{gb}\left(k,\eta_0\right) &\simeq & -\frac{\Psi(k,\eta_{dec})}{6}
\left( k\eta_0\right)^2-D_{gb}^{(in)}ks\sin\left( ks\right)  \nonumber\\
&&- \Psi _0\left[g\left(k,\eta_{dec}\right)+\eta_{dec}\dot{g}\left(k,\eta_{dec}\right)\right] \;, 
\label{eq:Db_t0}
\end{eqnarray}
The dominant term here is $\frac{\Psi(k,\eta_{dec})}{6}\left( k\eta_0\right)^2 \simeq 
\Psi_0(\eta_0/\eta_{eq})^2$ which comes from the baryons falling into the gravitational 
potential of dark matter. In addition, we have a growing function oscillating like 
$ks\sin\left( ks\right)$ plus a correction due to the $\Psi_0$-term which  slightly affects 
the period of the oscillations, see discussion below Eq.~(\ref{eq:DbSmall}).

To better understand the expression in square brackets, let us consider Equation (\ref{gx}), 
which for $ks \gtrsim \pi$ shows that
\begin{eqnarray}
g(k,\eta_{\rm dec}) + \eta_{\rm dec} \dot{g}(k,\eta_{\rm dec}) &\simeq & \eta_{\rm dec} \dot{g}(k,\eta_{\rm dec}) \nonumber \\
&=& -18\, \pi ^{-1} ks \cos(ks) \;,
\end{eqnarray}
This suggests that the position of the troughs and peaks in the matter 
power spectrum may  differ slightly  from those of $ks\sin\left(ks\right)$ 
and this difference is proportional to $\Psi_0$ which is in turn proportional 
to $\Omega_mh^2$.
Since the pre-factor is small, we can approximate $ks\sin(ks) 
+ \ep ks\cos(ks)$ by $ks\sin(ks + \ep) + {\cal O}(\ep^2)$.
With this, we expect that the positions of the troughs and the 
peaks in the matter power spectrum are approximately given by
\begin{equation}\label{eq:k_fit}
k_n = \frac{n\pi}{2s}\left(1+\beta _n\cdot\Omega _mh^2\right) \;,
\end{equation}
where $n=3,7,11,\ldots$ for the troughs and $n=5,9,13,\ldots$ for the peaks, and where 
$\beta _n $ is a parameter that takes into account the correction which affects 
mainly the lowest $k_n$'s and which can be fitted by comparing with numerical results.
Since $\Psi_0 \propto -D_g$ and $g +\eta_{\rm dec}\dot g <0$, we naively expect
$\ep>0$ and hence $\beta_n<0$. We shall see, however, that the approximations made for
the decoupling redshift actually lead to  $\beta_n>0$.

\section{Fit of the acoustic peak positions}
\label{s:1pk}

Eq.~(\ref{eq:k_fit}) allows us to localize the troughs and the peaks in the matter power 
spectrum. We finally want to derive an explicit form for the sound horizon at decoupling 
$s$, defined as the comoving distance that a wave can travel prior to decoupling $t_{dec}$:
\be\label{eq:s_int}
 s \equiv  \int _0^{t_{dec}}c^{(\gamma b)}_s\left( 1+z\right) dt ' \;.
\ee
The sound speed of the photon-baryon plasma is given in appendix \ref{app:TCA}, 
Eq.~(\ref{eq:c2s_gr}).

This integral can be computed exactly if we neglect the contribution of dark energy to $z$,
which is a very good approximation for the redshifts we are interested in.
The subscripts $b$, $c$ and $m$ refers to baryons, CDM and non-relativistic matter (baryons plus CDM), respectively; we define the density parameter $\om_X\equiv\Om_Xh^2$ for the species $X$.
The subscript $\ga$ refers to photons while the subscript $r$ refers to the
density in relativistic particles at the time of equal matter and radiation, which probably 
also comprises three types of neutrinos.
We consider, $\om_c$, $h$, $\om _m$ as independent cosmological parameters, keeping 
the first two fixed and varying the latter. We then write the remaining parameter as 
$\omega _b = \omega _m - \omega _c $.
This yields
\begin{eqnarray} 
s &\simeq& 
  \frac{h}{H_0\sqrt{3}} \int_{1+z_{dec}}^\infty\frac{dx}{x\sqrt{(x+r)
    (x\omega_r +\omega_m )}} \nonumber \\ 
  &=& \frac{4h}{3H_0}\sqrt{\frac{\om_\ga}{\om_b\omega_m}}\times  \nonumber \\
 && \  \log\left(\frac{\sqrt{1+\frac{r}{1+z_{dec} }} +\sqrt{\frac{
r\omega_r}{\omega_m} +\frac{r}{1+z_{dec} }}}{1 +
 \sqrt{\frac{r\omega_r}{\omega_m}}}\right)  .
\label{eq:s_develop}
\end{eqnarray}
$H_0$ is the value of the Hubble parameter today and $r=(1+z)R=3\om_b/(4\om_\ga)$ is the $r$-parameter defined in~\cite{Vonlanthen:2010cd}.
It is worth stressing that here $z_{dec}$ denotes the baryon decoupling redshift. 
In fact, even if radiation decoupling epoch is $z_{*}\simeq 1100$, the scattering rate of electrons is sufficient to keep the matter temperature $T_m$ equal to radiation temperature $T_{\gamma}$ down to redshifts $z\sim 100$.
Then, to extract BAO's information is necessary to integrate up to the baryon decoupling epoch, also known as drag epoch.
However, for $z<z_{*}$ Equation (\ref{eq:c2s_gr}), which gives $c^{(\gamma b)}_s$ during the tight coupling regime, no longer holds.
Nevertheless, we note that in the {\sc Recfast} code $T_m$ is set to $T_{\gamma}$ (plus a small correction) until a typical switching redshift $z\sim 850$, after which the full evolution equation for $T_m$ are considered \cite{Scott:2009}.
We find convenient for the fit to artificially shift the baryon decoupling redshift to the {\sc Recfast} `switch' value, $z_{dec}=850$, still considering Equation (\ref{eq:c2s_gr}) for $c^{(\gamma b)}_s$.
Furthermore, the $z_{dec}$ value is almost independent of the cosmological parameters within the range of our interest\cite{mukhanov}.

This approximate $z_{dec}$ treatment force us to reconsider the interpretation of Equation (\ref{eq:k_fit}).
In fact, we extrapolate this formula to correct also the uncertainties related to $z_{dec}$.
As we will see, this implies in particular that $\beta_{n} > 0$ in our fit.

\begin{figure}[h]
\begin{center}
\includegraphics[width=\columnwidth]{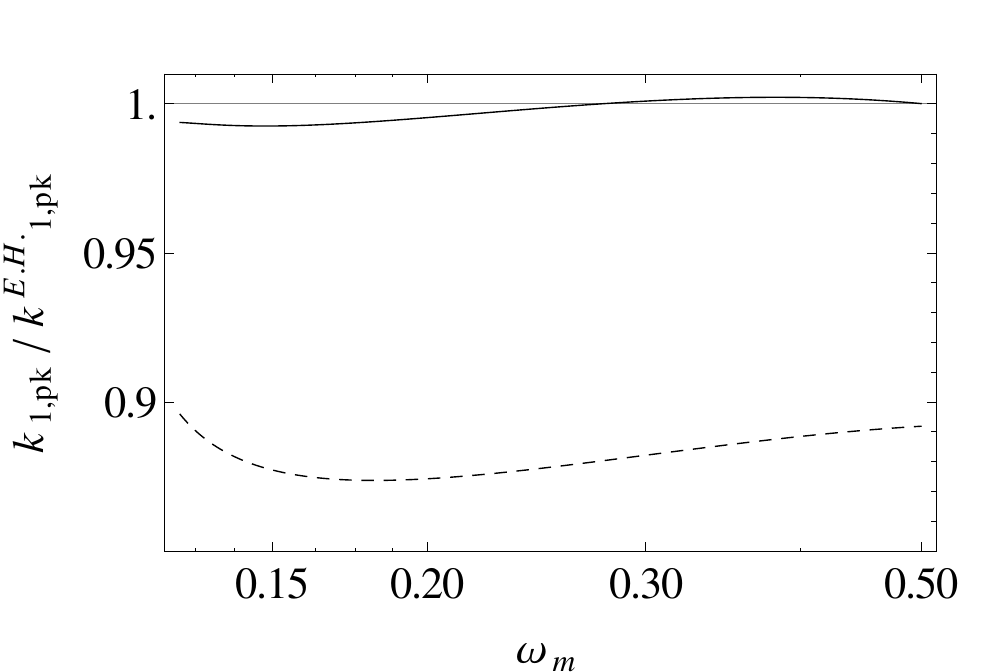}
\end{center}
\caption{Comparison of first peak position for $h=0.70$, $\omega_m=0.13$, $\omega_b=0.02$.
$k_{1,pk}$ is evaluated by using Eq. (\ref{eq:s_develop}) for $s$ and the $z_{dec}$ employed in \cite{EH98} (solid line) and in {\sc Recfast} (dashed).
$k_{1,pk}^{E.H.}$ is the fit proposed
in~\cite{EH98}.}
\label{k_kEH}
\end{figure}

In \cite{EH98}  a fitting formula for the matter transfer function of a CDM plus baryon 
Universe can be found. The curvature and also the cosmological constant are 
neglected. Since the latter do not contribute significantly to the sound horizon at 
decoupling, this approximation is still valid in a $\La$CDM Universe.
In Figure \ref{k_kEH} we compare the first peak positions $k_{1,pk} / k_{1,pk}^{E.H.}$ 
evaluated  approximatively as $5\pi / 2s$.
The wavenumber $k_{1,pk}$ is calculated by using Eq.~(\ref{eq:s_develop}) with $z_{dec}$ in accordance to {\sc Recfast}, while 
$k_{1,pk}^{E.H.}$ is calculated according to the sound horizon at decoupling employed 
in Eisenstein and Hu (1998) \cite{EH98}.
The parametric formulas for $s$ lead to an agreement within $1\%$ for the first peak position when considering the same recombination redshift as used in \cite{EH98}.
Instead, if we use the value consistent with {\sc Recfast} in Equation (\ref{eq:s_develop}), we find that in \cite{EH98} the position is systematically overestimated (in terms of $k$) by about $10\%$.
Furthermore, 
the full fitting formula proposed in \cite{EH98}, accounting also for the $\Om_m$-correction, 
yields disagreements up to $15\%$ with respect to the numerical results obtained with {\sc Camb}.

\subsection{Fit of the acoustic peak positions}
\begin{figure}[h]
\begin{center}
\includegraphics[width=\columnwidth]{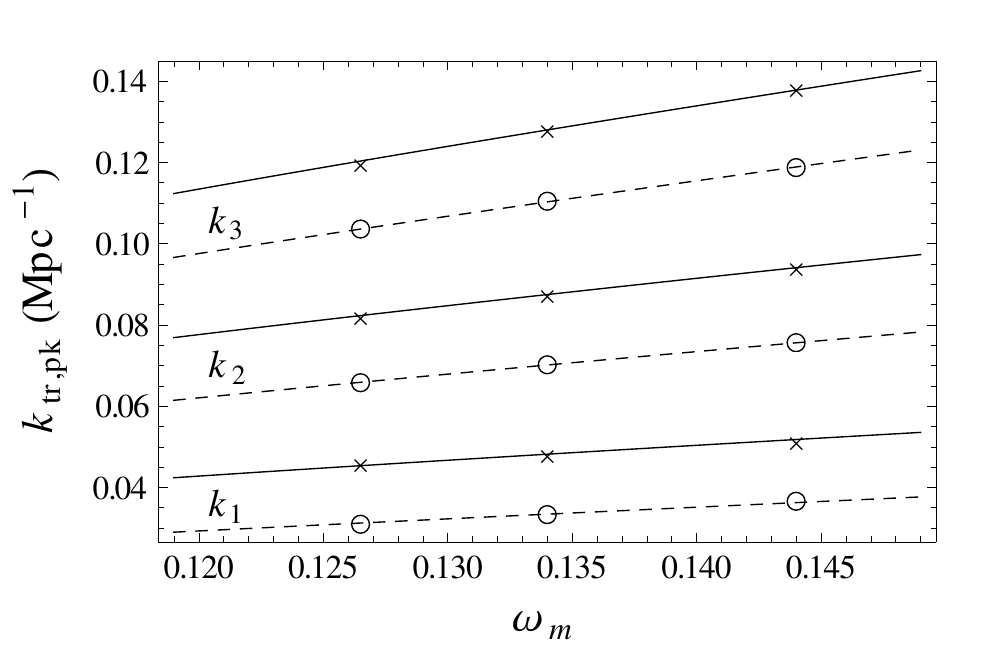}
\end{center}
\caption{First three troughs and peaks fit. The circles and the crosses are the numerical data for troughs and peaks respectively. The dashed and solid lines are the fits for troughs and peaks respectively.
We fixed $h=0.70$, $\omega_c=0.114$, $\omega _{r}=4.17\cdot 10^{-5}$, $\omega _{\gamma}=2.48\cdot 10^{-5}$ and $H_0^{-1} = 2997.9\ h^{-1}\ \text{Mpc}$.}
\label{fig:Fit_k123}
\end{figure}

Let us discuss, for illustrative purpose, the fit of the first three troughs and peaks in the matter power spectrum, Figure \ref{fig:Fit_k123}.
As we are not now interested in precision, the fit is compared to a numerical code \citep{FM} that agrees with {\sc Camb} within about $5\%$.

It is clear that the form of the fit is adequate to reproduce the numerical results, but let us 
consider Table \ref{tab:B_123}, which reports the fit parameters obtained, to check 
our expectations.
\begin{table}[h]
\centering
\begin{tabular}{|l|cc|}
  \hline
  Order & Troughs & Peaks \\
  \hline
  $1st$  & 0.25  & 0.07\\
  $2nd$ & 0.12  & 0.08\\
  $3rd$ & 0.12  & 0.10\\
  \hline
\end{tabular}
     \caption{\label{tab:B_123} $\beta$-correction, defined as $\beta_n\omega_m$. We have fixed $\omega_m=0.144$.}
\end{table}
From Equation (\ref{eq:k_fit}) we see that the relative importance of the $\beta$-correction 
is given by $\beta_n\omega_m$.
As explained above, this correction is due to the fact that the first nodes of the transfer function slightly differ from those of $\sin(ks)$, also for 
$ks\gtrsim\pi$, because of the velocity overshoot effect;
together with the approximation for $z_{dec}$ which is too high, hence an $s$ which is 
too small,  leads to $\beta_n > 0$.
Indeed, as shown in Table \ref{tab:B_123} a correction of about $25\%$ is obtained 
for the first trough.
This $\beta$-correction is larger than the other cases, for which is about $10\%$, and it is due to the fact that the corresponding $ks= 3\pi/2$ is the closest to the critical value $ks=\pi$.
Actually, we also note that the corrections for the troughs are larger than for the peaks; this 
is due to the method used to extrapolate the trough and peak positions, but here we neglect 
this detail.

\subsection{The fitting formula for the first peak}
The first peak position in the matter power spectrum is conveniently fitted by
\begin{subequations}
\begin{equation} \label{eq:k1pk_final}
k_{1,pk} = \frac{5\pi}{2s}\left(1+0.276\;\Omega _mh^2\right) \; .
\end{equation}
Inserting the wellknown photon density $\om_\ga$ and $H_0/h$ we obtain for $s$ 
\begin{equation} \label{eq:s_final}
s = 19.9\ \left(\omega _m\omega _b\right)^{-1/2} \log{\left[ U\left( \om _b,\om _m\right)\right] } 
\text{ Mpc} \;,
\end{equation}
and
\begin{equation*}
U\left( \om _b,\om _m\right) = \frac{1.12\sqrt{\frac{\omega _b}{\omega _m}\left(1+28.18\ \omega _m\right)}+\sqrt{1+35.54\ \omega _b}}{1+1.12\sqrt{\frac{\omega _b}{\omega _m}}} \; .
\end{equation*}
\end{subequations}
Note that the units are Mpc, not $h^{-1}{\rm Mpc}$.
With this, the  fit  (\ref{eq:k1pk_final}) deviates by less than $5\%$ from the 
numerical results of {\sc Camb} for the range of cosmological parameters, around the values reported in \cite{spergel}, $0.70\lesssim h\lesssim 0.75$, $0.100\lesssim \omega _c\lesssim 0.130$ and $0.0125\lesssim \omega _b\lesssim 0.030$.

\begin{figure}[h]
\begin{center}
\includegraphics[width=\columnwidth]{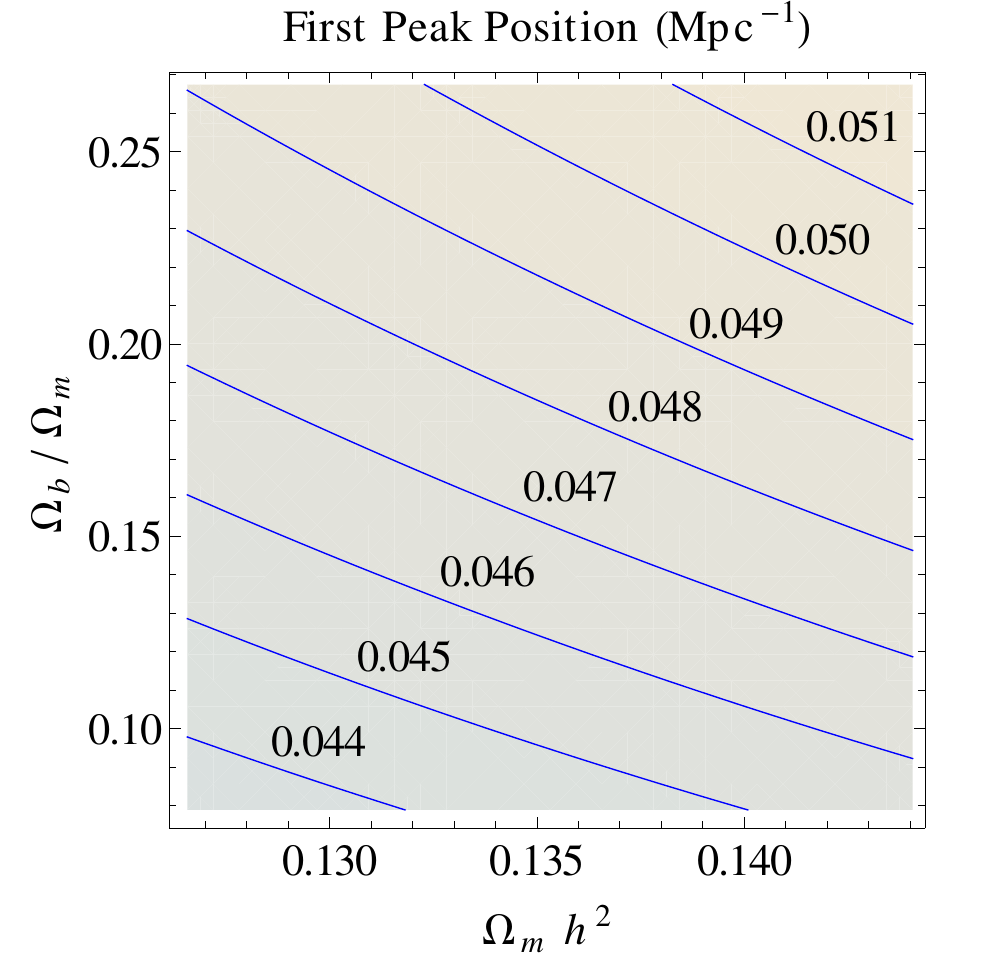}
\end{center}
\caption{The location of the first peak in $\text{Mpc}^{-1}$ as a function of baryon and matter density parameters. Lines of constant $k_{1,pk}$ are indicated.}
\label{fig:Contour_Om_ObOm}
\end{figure}

In Figure \ref{fig:Contour_Om_ObOm} we show the location of the first peak $k_{1,pk}$ 
as a function of baryon and matter density parameters.
As the baryon fraction $\Omega_b/\Omega_m$ increases, the first peak is 
shifted to smaller scales, since the sound speed and with it $s$ decrease. 
The value of $k_{1,pk}$ also increases with $\Omega_m$, due to the larger contribution of 
the $\Omega_m$-term in Equation (\ref{eq:k1pk_final}).

\section{Conclusions}
\label{s:concl}

Matching the tight coupling approximation, Eq.~(\ref{eq:DbSmall}), to the solutions after decoupling, allowed us to develop further the approach initiated in~\cite{HS1996}.
This yields an analytical formula for the location of the peaks 
 and troughs in the matter power spectrum, Eq.~(\ref{eq:k_fit}).  The formula has 
 the same form as the one given in~\cite{EH98}.

Using the same approximation for the recombination history as~\cite{EH98}, we obtained results compatible with~ \cite{EH98} within about $1\%$,  even though we consider very different cosmological parameters, $\Om_\La \sim 0.7$ as compared to $\Om_\La\sim 0$ which was considered in~\cite{EH98}.
This shows that the acoustic peak positions are not really 
sensitive to $\Om_\La$ but only to $\om_m$,  $\om_b$ and of course $\Om_{\rm total}$.
This corresponds also to the findings of~\cite{Vonlanthen:2010cd}.
However, considering an improved recombination history, i.e., using {\sc Recfast}, the fit proposed in~\cite{EH98} for the location of the first peak in the matter power spectrum no longer holds.

This leads us to propose an improved fitting formula for the position of the first peak 
obtained by running {\sc Camb}, see equations (\ref{eq:k1pk_final}), (\ref{eq:s_final}) and Figure \ref{fig:Contour_Om_ObOm}.
The fit yields the location of the first peak in a convenient range of cosmological parameters around the values reported in \cite{spergel}, with an accuracy of about $5\%$ with respect to the numerical results of {\sc Camb}, whereas the fitting proposed in \cite{EH98} disagrees by up to $15\%$ with {\sc Camb} within the range of parameters we explored.
The $5\%$ error estimation takes into account not only discrepancies with respect to {\sc Camb}, but also minor contributions that may come, e.g., from neglected non-linear effects, or variations of $1\%$ order in the ionization history.
The latter might be improved by using a different recombination code like {\sc Rico} and, 
as minor contribution, by improving treatment of reionization .

The fit may be improved on the one hand by better approximations of the baryon density 
$D_{bg}$ and  by improving the recombination history and on the other hand
by improving the fitting formula (\ref{eq:k1pk_final})-(\ref{eq:s_final}) itself.
This may also imply a more accurate discussion of the baryon decoupling redshift, see Equation (\ref{eq:s_develop}), which here has been artificially shifted to $z_{dec}=850$, based on 
the  implementation of {\sc Recfast}, independently of cosmological parameters.
However, we expect that most of the individual effects will lead at best to a 1\% improvement and recalling that on this level e.g. also the value of the scalar spectral index affects the peak position, we believe that a better fitting formula would be very complicated and to obtain significant improvements in the evaluation of the first peak position one has to resort numerical calculations which might need to go to
second order in perturbation theory.

Our fitting formula is especially useful for a first estimate of the effects of changing cosmological parameters on the positions of the baryon acoustic peaks.
\vspace{1cm}\\
{\bf Acknowledgments}\\
We thank Gianfranco de Zotti and Sabino Matarrese for discussions. 
This work is supported by the Swiss  National Science Foundation.


\appendix
\section{Tight coupling approximation}
\label{app:TCA}
In order to derive Eq.~(\ref{eq:bar_an}), we consider the evolution of baryon perturbations  
during the tight coupling regime. We follow~\cite{dodelson} and~\citep{durrer}, where 
the  evolution of photon perturbations is discussed in detail.
Since baryons are coupled via Thomson scattering to photons, the evolution of baryon perturbations is related to that of photons by the differential optical depth $\dot{\kappa}=a \si_T n_e$, where $\si_T$ denotes the Thomson cross section and $n_e$ the electron number density.
Indeed, the equations governing the baryon perturbations evolution read~\citep{durrer}:
\begin{subequations}
\begin{eqnarray}
&&\dot{D}_{gb} = -kV_b \;, \label{eq:dDb}\\
&&\dot{V}_b + \mathcal{H}V_b = k\Psi + \frac{\dot{\kappa}}{R}\left(V_{\gamma}-V_b\right) 
\;, \label{eq:dVb}
\end{eqnarray}
where we use $R=\frac{3\rho _b}{4\rho _{\gamma}}$ and $V$ denotes the veolocity perturbation.
We also write the first moments of the Boltzmann equation for photons
\begin{eqnarray}
&&\dot{D}_{g\gamma} = -\frac{4}{3}kV_{\gamma} \;, \label{eq:dDg}\\
&&\dot{V}_{\gamma} = 2k\Psi + \frac{1}{4}kD_{g\gamma} - \dot{\kappa}\left(V_{\gamma}-
  V_b\right) \;. \label{eq:dVg}
\end{eqnarray}
\end{subequations}
Since CDM does not interact other than gravitationally, we do not need to consider its
evolution here. 

If we take the limit $\dot{\kappa}\rightarrow\infty$ in Equation (\ref{eq:dVg}) we find
\begin{equation} \label{eq:Vb_Vg_TC}
V_b=V_{\gamma} \;.
\end{equation}
This zero-order tight coupling solution leads to an important consideration: during the tight coupling phase, perturbations between baryons and photons are roughly adiabatic on all scales due to Thomson scattering. 

Using this zero-order result (in $1/\dot{\kappa}$) back in the l.h.s. of Equation (\ref{eq:dVg}) we find the leading order equation:
\begin{eqnarray}
V_{\gamma} - V_b = \frac{k}{\dot{\kappa}}\left(2\Psi + \frac{1}{4}D_{g\gamma}\right) - 
\frac{1}{\dot{\kappa}}\dot{V}_b \;.
\end{eqnarray}
Using this in Eq. (\ref{eq:dVb}) we obtain:
\begin{eqnarray}\label{eq:dVb_2}
\dot{V}_b + \frac{R}{1+R}\mathcal{H}V_b - \frac{k}{4\left(1+R\right)}D_{g\gamma} = \frac{2+R}{1+R}k\Psi \;.
\end{eqnarray}
Differentiating Eq. (\ref{eq:dDb}) and using Eq. (\ref{eq:dVb_2}) to replace $\dot{V}_b$ we find:
$$
\ddot{D}_{gb} = \frac{R}{1+R}\mathcal{H}kV_b - \frac{k^2}{4(1+R)}D_{g\gamma} - \frac{2+R}{1+R}k^2\Psi \;.
$$
We use again Eq. (\ref{eq:dDb}) to substitute $kV_b$ and the fact that until photons and 
baryons are tightly coupled, the adiabaticity condition $D_{g\gamma}=\frac{4}{3}D_{gb}$ 
holds \cite{durrer}. Then, we have	
$$
\ddot{D}_{gb} = -\frac{R}{1+R}\mathcal{H}\dot{D}_{gb} - \frac{k^2}{3\left(1+R\right)}D_{gb} -  \frac{2+R}{1+R}k^2\Psi \;.
$$
Using $R \propto a$, the comoving Hubble parameter writes $\mathcal{H} = \dot{a} / a = \dot{R} / R$.
We write the sound speed of the photons plus baryons system as
\begin{equation} \label{eq:c2s_gr}
c^{(\gamma b)}_s \equiv \sqrt{\frac{\dot{P}_{\gamma}}{\dot{\rho}_{\gamma}+\dot{\rho}_b} } = 
\frac{1}{\sqrt{3\left(1+R\right)}} \;,
\end{equation}
where we also used $P_b=0$ and $P_{\gamma}=\rho_{\gamma}/3$ for the baryon and photon pressure, respectively.

We finally write the equation for the baryon density perturbations as
\begin{eqnarray} \label{eq:HarmOscil}
\ddot{D}_{gb}  + \frac{\dot{R}}{1+R}\dot{D}_{gb} + k^2\left(c^{(\gamma b)}_s\right)^2D_{gb} 
= F(k,t) \;,
\end{eqnarray}
where we have defined the forcing function
\begin{equation}
F(k,t) = - \frac{2+R}{1+R}k^2\Psi(k,t) \;.
\end{equation}
Eq.~(\ref{eq:HarmOscil}) represents damped, driven oscillations of the baryon density 
perturbation. The second term on the left-hand side is the damping of oscillations due 
to the expansion of the universe.
The third term on the left-hand side is the restoring force due to the pressure.
The forcing function is governed by the gravitational potential perturbations.
These oscillations are called `acoustic oscillations' since, as in acoustic waves, the photon-baryon fluid cannot simply collapse under gravity because of the restoring force provided by the pressure which leads to oscillations.

To obtain an analytical solution to Eq.~(\ref{eq:HarmOscil}), we first find the solutions to the homogeneous equation through the WKB approximation \cite{HS1995}, valid for slow varying $R$,  inside the sound horizon at decoupling given by $s\simeq c_s^{(\gamma b)}\eta_{dec}$. Then we obtain a particular solution by the standard Green's method imposing adiabatic initial conditions.
This yields the general tight coupling solution for the baryon density perturbation
\begin{eqnarray}
D_{gb}^{(t.c.)} \left(k, \eta\right) &=& D_{gb}^{(in)} \left(\frac{1}{1+R\left(\eta\right)}\right)^{1/4} \cos\left( kr_s\right) \nonumber \\
&& - E\left( k,\eta \right) \;,
\end{eqnarray}
where
\begin{eqnarray*}
E\left( k,\eta \right) &=& \left( 1+R\left(\eta\right)\right)^{-1/4}  
\int _0^{ \eta}d\zeta\ \left[ \frac{2+R\left( \zeta \right)}{\left(1+R\left(\zeta\right)\right)^{3/4}} \right. \nonumber \\
&& \times \left. \frac{\sin\left[kr_s\left( \eta\right)-kr_s\left( \zeta \right)\right]}{kc^{\left( \gamma b \right)}_s\left(\zeta\right)} k^2\Psi\left( k,\zeta\right) \right] \;.
\end{eqnarray*}

\bibliographystyle{h-physrev}
\bibliography{biblio}

\end{document}